\documentclass[12pt]{iopart}
\usepackage{iopams}
\usepackage{graphicx}
\usepackage{hyperref}
\usepackage{xcolor}

\newcommand{\bbr}{{\mathbf{r}}}
\newcommand{\R}{\mathbb{R}}
\newcommand{\Z}{\mathbb{Z}}
\newcommand{\cl}{{\rm cl\,}}
\newcommand{\Int}{{\rm int\,}}

\newtheorem{thm}{Theorem}
\newtheorem{remark}{Remark}

\begin{document}

\title[Reachable sets for two-level open quantum systems]{Reachable sets for two-level open quantum systems driven by coherent and incoherent controls}
\author{Lev Lokutsievskiy$^{1,2}$ and Alexander Pechen$^{1,3}$\footnote{Author to whom any correspondence should be addressed.}}
\address{$^1$ Steklov Mathematical Institute of Russian Academy of Sciences, 8 Gubkina str., Moscow 119991, Russia}
\address{$^2$ Lomonosov Moscow State University, GSP-1, Leninskie Gory, Moscow 119991, Russia}
\address{$^3$ National University of Science and Technology ''MISIS'', 4 Leninsky Prosp., Moscow 119991, Russia}
\eads{lion.lokut@gmail.com and apechen@gmail.com}

\begin{abstract}
In this work, we study controllability in the set of all density matrices for a two-level open quantum system driven by coherent and incoherent controls. In [A. Pechen, Phys. Rev. A {\bf 84}, 042106 (2011)] an approximate controllability, i.e., controllability with some precision, was shown for generic $N$-level open quantum systems driven by coherent and incoherent controls. However, the explicit formulation of this property, including the behavior of this precision as a function of transition frequencies and decoherence rates of the system, was not known. The present work provides a rigorous analytical study of reachable sets for two-level open quantum systems. First, it is shown that for $N=2$ the presence of incoherent control does not affect the reachable set (while incoherent control may affect the time necessary to reach particular state). Second, the reachable set in the Bloch ball is described and it is shown that already just for one coherent control any point in the Bloch ball can be achieved with precision $\delta\sim \gamma/\omega$, where $\gamma$ is the decoherence rate and $\omega$ is the transition frequency. Typical values are $\delta\lesssim10^{-3}$ that implies high accuracy of achieving any density matrix. Moreover, we show that most points in the Bloch ball can be exactly reached, except of two lacunae of size $\sim\delta$. For two coherent controls, the system is shown to be completely controllable in the set of all density matrices. Third, the reachable set as a function of the final time is found and shown to exhibit a non-trivial structure.
\end{abstract}

\section{Introduction}

Quantum control, that is control of individual quantum systems, in an important tool necessary for development of modern quantum technologies~\cite{Glaser2015,ButkovskyBook1994,ShapiroBrumerBook2003,TannorBook2007,AlessandroBook2007,LetokhovBook2007,FradkovBook2007,BrifChakrabartiRabitz2010,WisemanMilburnBook2010,DongPetersen2010,Moore2011,ZagoskinBook2011,CPKoch2016OpenQS,Stefanatos2020}. Often in experiments controlled quantum systems are interacting with the environment, that is, they are open quantum systems. This circumstance motivates the development of efficient methods for controlling open quantum systems. 

The environment is sometimes considered as an obstacle having deleterious effects on the dynamics of the controlled system. However, the environment can also be used for controlling quantum systems via its temperature, pressure, or more generally, non-equilibrium spectral density. A general method of incoherent control using this spectral density of the environment, including also in combination with coherent control, either subsequent of simultaneous, was developed and studied for any multilevel quantum systems in~\cite{PechenRabitz2006}. In this method, spectral density of the environment, i.e., distribution of its particles in their momenta and internal degrees of freedom, is used as a control function to drive the system. This spectral density in some cases can be  thermal (i.e., Planck distribution), but in general it can be any non-equilibrium non-negative function, even possibly depending on time, of momenta and internal degrees of freedom of environmental particles. Its non-negativity follows from its physical meaning as density of particles. A natural example of such environment is the environment formed by incoherent photons. For incoherent photons, the control is performed by realizing various non-equilibrium distribution functions $n_{\omega,\alpha}(t)$ (for photons $\omega$ is frequency and $\alpha$ is polarization). Such control by time dependent temperature can be realized experimentally. For example, implementing fast and controlled temperature variations for non-equilibrium control of thermal and mechanical changes in a levitated system is provided in~\cite{arXiv:2103.10898}. Another approach to incoherent control is to use back-action of non-selective quantum measurements to manipulate the quantum system, as was proposed in~\cite{PechenIlinShuangRabitz2006} and studied, e.g., in~\cite{ShuangPechenHoRabitz2007,ShuangZhouPechenWuShirRabitz2008,PechenTrushechkin2015}.

Given a controlled system, establishing the degree of its controllability is among the most important practical questions. For closed quantum systems, a detailed analysis of controllability for various cases was performed~\cite{Huang1983,Tarn1984,Turinici2001,Albertini2001,Fu2001,Schirmer2001,Schirmer2002,Altafini2002,Polack2009,Boscain2015}. Controllability for coherently controlled open systems with GKSL master equations was investigated in~\cite{Altafini1,Altafini2}. Time-optimal control of dissipative two-level open quantum systems was studied using geometric control theory and other methods in~\cite{SugnyPRA2007,BonnardSIAM2009,BonnardIEEE2009,StefanatosPRA2009,StefanatosSCL2010,MorzhinIJTP2019,MorzhinLJM2020,MorzhinAIP2021}. Local properties such as the absence of traps for controlling a qubit have also been proved~\cite{PechenPRA2012,PechenJPA2017,VolkovJPA2021}.

Initially for incoherent control it was not clear to what degree it allows for controlling quantum systems. In~\cite{Pechen2011}, it was shown that combination of coherent and incoherent controls allows to approximately steer {\it any} initial density matrix to {\it any} given target density matrix for a generic quantum  system. This property approximately realizes complete controllability of open quantum systems in the set of all density matrices --- the strongest possible degree of quantum state control. The proposed scheme has several important features. (1) It was obtained with a physical class of Gorini-Kossakowsky-Sudarchhan-Lindblad (GKSL) master equations well known in quantum optics and derived in the weak coupling limit. (2) It was obtained for almost all values of parameters of this class of master equations and for multi-level quantum systems of arbitrary dimension. (3) For incoherent controls an explicit analytic solution (not numerical) was obtained. (4) The scheme is robust to variations of the initial state --- the optimal control steers simultaneously {\it all} initial states into the target state, thereby physically realizing all-to-one Kraus maps previously theoretically exploited for quantum control in~\cite{Wu2007}.

However, estimate of the precision to which one can steer an initial state to a target state as a function of the system parameters (i.e., transition frequencies and decoherence rates) was not found in~\cite{Pechen2011}. The present work provides a detailed rigorous analytical study of reachable sets for two-level open quantum systems driven by coherent and incoherent control which fills this gap. Situations with one and two coherent controls are considered. First, it is shown that for $n=2$ level quantum systems the presence of incoherent control does not affect the reachable set, while it may affect the time necessary to reach particular state.  Second, an explicit description of the reachable set in the Bloch ball is provided. It is shown that already for one coherent control all points in the Bloch ball can be exactly reached, except of points in two lacunae of size $\delta\sim\gamma/\omega$ around two pure states, where $\gamma$ is the decoherence rate and $\omega$ is the transition frequency. Thus, on one hand, it is shown that any point in the Bloch ball can be achieved with precision $\delta\sim\gamma/\omega$. On another hand, it is shown that any pure state (except for the two trivial ones) can not be achieved with precision better than $\delta\sim \gamma/\omega$. Typical values are $\delta\lesssim10^{-3}$ that implies high accuracy of achieving any density matrix. Third, a numerical description of the reachable set as a function of the final time is provided and it is found that this time evolution of the reachable set exhibits a non-trivial structure.

Controllability problems for $n$-level coherently controlled open systems with GKSL master equations was investigated in~\cite{Altafini1}. The two-level case with three coherent controls driving rotations around all three axes on the Bloch sphere was considered in~\cite{Altafini2}, where controllability was established for amplitude damping master equation (with affine GKSL superoperator). In this work, we analyze controllability for two-level open quantum systems driven by incoherent control and either one and two coherent control fields. In the case of one control we find two unreachable lacunae of small size, whereas for two controls the system becomes completely controllable. The case with one coherent control is the most typical. For this case, if the two-level quantum system is closed (i.e., not interacting with the environment, so that $\gamma=0$) it would be completely controllable in the set of pure states. As we show, even if the quantum system is open, then just with only one coherent control it is controllable with a high degree of precision and with two controls it is controllable exactly in the set of {\it all} density matrices.

The structure of the paper is the following. Formulations of the control problem in terms of GKSL master equation and in terms of Bloch vector parametrization are provided in Sec.~\ref{sec:Bloch}. In Sec.~\ref{sec:coherent_and_incoherent_controls} it is shown that incoherent control for a two-level quantum system does not affect the reachable set. Asymptotically reachable points are described in Sec.~\ref{sec:asymptotically}. Size of the reachable set and $\delta$ are estimated in Sec.~\ref{sec:size}. In Sec.~\ref{sec:size} we also numerically compute the reachable sets for several different final times and even a full movie showing how the reachable set grows with increasing the final time. The numerical computations are done using the fast method for numerical engineering of optimal coherent control designed in Sec.~\ref{sec:fast}. Conclusions section~\ref{sec:conclusions} summarizes the results.

\section{Parametrization by Bloch vector}\label{sec:Bloch}

In this work we study analytically controllability of a two-level quantum system (for shortness we call it as a qubit) coupled to an environment. Density matrix $\rho$ of a qubit is a positive trace one $2\times 2$ matrix, $\rho\in\mathbb C^{2\times 2}$, $\rho\ge 0$, $\Tr\rho=1$. The qubit is driven by two types of control: coherent and incoherent. Its density matrix satisfies the following master equation~\cite{PechenRabitz2006}
\begin{eqnarray}\label{f1}
\fl\frac{\rmd \rho(t)}{\rmd t} =
-\frac{\rmi}{\hbar}\Bigl[\omega \sigma_z + \kappa u(t)\sigma_x, \rho(t)\Bigr] 
+\gamma (n(t)+1) \Big( \sigma^- \rho(t) \sigma^+ 
- \frac{1}{2} \Big\{ \sigma^+ \sigma^-, \rho(t) \Big\} \Big) \nonumber \\
+\gamma n(t)\Big(  \sigma^+ \rho(t) \sigma^-   
- \frac{1}{2} \Big\{ \sigma^- \sigma^+, \rho(t) \Big\} \Big).
\end{eqnarray}
Here $u(t)\in\R$ is the coherent control, $n(t)\in\R_+$ is the incoherent control, $\omega>0$ is the transition frequency of the qubit, $\kappa>0$ is its coupling to the coherent control (e.g., dipole moment), $\gamma\ge 0$ is the decoherence rate (for a given quantum system $\omega$, $\kappa$, and $\gamma$ are some constants), $\sigma_x$ and $\sigma_z$ are the $X$ and $Z$ Pauli matrices, $\sigma^+=\left(\begin{array}{cc}
0 & 0 \\
0 & 1  
\end{array}\right)$ and $
\sigma^-= \left(\begin{array}{cc}
0 & 1 \\
1 & 0
\end{array}\right)$ are the raising and lowering matrices, $[\cdot,\cdot]$ and $\{\cdot,\cdot\}$ stand for commutator and anti-commutator of two matrices, respectively. The Hamiltonian term in the commutator is written without loss of the generality. We set in the rest Planck constant $\hbar=1$.

Since the system is affine in both controls $u(t)$ and $n(t)$, we assume as usual that $u\in L^1$ and $n\in L^1$.

Density matrix of the qubit can be conveniently parametrized as
\[
\rho=\frac{1}{2}\left(\mathbb I+r_x\sigma_x+r_y\sigma_y+r_z\sigma_z\right),
\]
where $\mathbf{r}=(r_x,r_y,r_z)\in\R^3$ and $(\sigma_x,\sigma_y,\sigma_z)$ are the Pauli matrices. Vector $\mathbf{r}$ belongs to the Bloch ball, $|\mathbf{r}|\le 1$. Pure states satisfy $|\mathbf{r}|=1$ and belong to the Bloch sphere. Mixed states satisfy $|\mathbf{r}|<1$.

In this representation, the dynamics of the controlled system can be written as
\begin{equation}
\label{eq:main_control system}
	\dot \mathbf{r} = \omega f_0(\mathbf{r}) + 2\kappa f_1(\mathbf{r})u + \gamma f_2(\mathbf{r})n.
\end{equation}
Here (see~\cite{MorzhinIJTP2019})
\begin{eqnarray*}
	f_0(\mathbf{r})
	&=
    \left(\begin{array}{ccc}
		0&-1&0\\
		1&0&0\\
		0&0&0
	\end{array}\right)\mathbf{r}
	-
	\frac{\gamma}{\omega}
    \left(\begin{array}{ccc}
		\frac{1}{2} & 0 & 0\\
		0 & \frac12 & 0\\
		0 & 0 & 1
	\end{array}\right)\mathbf{r}
	+
	\frac{\gamma}{\omega}
    \left(\begin{array}{ccc}
		0\\0\\1
	\end{array}\right);\\
	f_1(\mathbf{r})&=\left(\begin{array}{ccc}
		0&0&0\\
		0&0&-1\\
		0&1&0
	\end{array}\right)\mathbf{r};
	\qquad
	f_2(\mathbf{r})=
	-\left(\begin{array}{ccc}
		\frac12 & 0 & 0\\
		0 & \frac12 & 0\\
		0 & 0 & 1
	\end{array}\right)\mathbf{r}.
\end{eqnarray*}

Below we study reachable sets for the system~(\ref{eq:main_control system}). It is known (see~\cite{Pechen2011}) that this system (and general $N$-level quantum control systems) is approximately controllable by both coherent and incoherent control. It means that there exists some small approximation accuracy $\delta>0$ such that for any two given end points $\bbr^0$ and $\bbr^1$ in the Bloch ball, there exists motion time $T\ge 0$ and controls $u(t)$ and $n(t)\ge 0$ for $t\in[0;T]$ such that the resulting trajectory starting at $\bbr(0)=\bbr^0$ ends at a point close enough to $\bbr^1$, so that $|\bbr(T)-\bbr^1|\le \delta$. 

In this paper, we analyze more precisely the sets of reachable points for the system~(\ref{eq:main_control system}). Recall that a point $\bbr^1$ is called {\it (exactly) reachable} from a point $\bbr^0$ if there exist $T\ge 0$ and controls $u(t)$ and $n(t)\ge 0$ for $t\in[0;T]$ such that the resulting trajectory starting at $\bbr(0)=\bbr^0$ ends exactly at $\bbr^1=\bbr(T)$. Similarly, a point $\bbr^1$ is called {\it asymptotically reachable} from $\bbr^0$ if there exist exactly reachable from $\bbr^0$ points that are arbitrarily close to $\bbr^1$.

Our main results are the following.
\begin{itemize}
	\item We find an exact estimation of the approximation accuracy $\delta$ and show that $\delta\sim \gamma/\omega$.
	\item We show that the reachable set for the system controlled only by coherent control (i.e.\ $n\equiv 0$) coincides up to some boundary points with the reachable set for the system controlled by both coherent and incoherent control.
	\item Obviously, if one leaves the system uncontrollable for a while (i.e. set $u=n=0$), then the state $\bbr(t)$ converges to $(0,0,1)$ exponentially fast. On one hand, we show that any point in the centered at origin ball of radius $1-\case{\pi}{4}\case{\gamma}{\omega}$ is exactly reachable from $(0,0,1)$. 
	\item On another hand, we demonstrate that any pure states (that is not a rotation of $(0,0,1)$ around axis $O\bbr_x$) can not be reached with accuracy better that $\delta\sim\gamma/\omega$. In particular, there are points in Bloch ball that cannot be exactly reached from the point $(0,0,1)$, and these points form a set of non-zero volume, which contains points on the distance $1-\alpha\gamma/\omega$ from the origin for any $\alpha$ such that $0\le\alpha<\frac12(1+\frac{\gamma^2}{\omega^2})^{-1/2}\sim\frac12$.
\end{itemize}

\section{Coherent and incoherent controls}
\label{sec:coherent_and_incoherent_controls}

In this section we prove the following surprising fact: for a two-level system incoherent control does not extend the set of reachable points. A point $\bbr^1$ is asymptotically reachable from $\bbr^0$ for a control system if and only if $\bbr^1$ belongs to the closure of the reachable set from $\bbr^0$. In other words, there exist trajectories of the system that start at $\bbr^0$ and end arbitrary close to $\bbr^1$.

Thus the main result of this section can be formulated as follows:

\begin{thm}
	For any starting point~$\bbr^0$ the sets of asymptotically reachable points for system~(\ref{eq:main_control system}) with or without incoherent control coincide.
\end{thm}

Now we introduce some useful notations to prove this statement. Denote by $B_+(\bbr^0)$ and $B_-(\bbr^0)$ the reachable sets of system~(\ref{eq:main_control system}) from the point $\bbr^0$ in forward and backward time, respectively. In other words, one is allowed to use \textit{both} coherent and incoherent control to reach points in $B_+(\bbr^0)$ and $B_-(\bbr^0)$. Also denote by $C_+(\bbr^0)$ and $C_-(\bbr^0)$ the reachable sets of system~(\ref{eq:main_control system}) from the point $\bbr^0$ in forward and backward time but for $n\equiv0$. In other words, one is allowed to use only \textit{coherent} control to reach points in $C_+(\bbr^0)$ and $C_-(\bbr^0)$. So, for example,
\[
\fl C_+(\bbr^0) = \big\{ \bbr^1 : \exists T\ge0,\bbr(t),u(t):\ \dot\bbr=\omega f_0(\bbr) + 2\kappa f_1(\bbr)u,\  \bbr(0)=\bbr^0,\bbr(T)=\bbr^1\big\},
\]
and sets $C_-(\bbr^0)$ and $B_\pm(\bbr^0)$ have similar definitions.

We are really interested only in sets $C^+(\bbr^0)$ and $B^+(\bbr^0)$, but sets $C^-(\bbr^0)$ and $B^-(\bbr^0)$ are also useful in our investigation. Obviously, using both controls, in general one can reach more points, i.e.\footnote{In this work, symbols $\subset$ and $\subseteq$ are used as  equivalent to denote not necessarily strict  inclusion of sets.} $B_+(\bbr^0)\supset C_+(\bbr^0)$.

Below we will prove that for any $\bbr^0$
\[
	\cl C_+(\bbr^0) = \cl B_+(\bbr^0),
\]
where $\cl X$ denotes closure of set $X$. Obviously $C_+(\bbr^0)\subset B_+(\bbr^0)$ and hence $\cl C_+(\bbr^0)\subset \cl B_+(\bbr^0)$. Therefore we need to demonstrate only the opposite inclusion. For this, consider the case when the incoherent control is switched off so that $n\equiv0$.

Denote by $\rme^{sf}$ the flow of a vector field $f(\bbr)$, i.e.\ $\rme^{sf}$ moves all point forward in time $s$ along solutions of the equation $\dot\bbr=f(\bbr)$.

We start with proving a simple, but very useful fact that for any point $\bbr^0$ and any $s$, the point $\rme^{sf_1}\bbr^0$ belongs to both sets $\cl C_\pm(\bbr^0)$.

First, note that $\rme^{sf_1}$ is the rotation around the first axis $O\bbr_x$ on the angle $s$. Hence, without loss of generality, one can assume that $s>0$ (if $s<0$, just put $\tilde s=s+2\pi k$ for some large enough~$k$). Choose an arbitrary small $\varepsilon>0$ and consider the controls $n\equiv 0$ and $u=1/(2\kappa\varepsilon)$ for $t\in[0;T]$ where $T=\varepsilon s$. In order to find the solution of the control system~(\ref{eq:main_control system}) with these controls, make time change $t=\varepsilon \tau$:
\[
\bbr'_\tau = \varepsilon \omega f_0(\bbr) + f_1(\bbr); 
\qquad \bbr(0)=\bbr^0;
\qquad \tau\in[0;s].
\]
Hence, the solution of the previous equation converges to the solution of the equation $\bbr'_\tau = f_1(\bbr)$ as $\varepsilon\to+0$. The end point at $\tau=s$ of the solution to the latter equation is $\rme^{sf_1}\bbr^0$ by definition. Hence, $\rme^{sf_1}\bbr^0\in \cl C_+(\bbr^0)$. For the set $C_-(\bbr^0)$,  similarly one can show $\rme^{sf_1}\bbr^0\in \cl C_-(\bbr^0)$.

Now we have a very powerful way of controlling the system: first apply very strong positive control $u=1/(2\kappa\varepsilon)$ for a very short time $T=\varepsilon s$ to reach (approximately) the point $\rme^{sf}\bbr^0$, then use any admissible control at $\rme^{sf}\bbr^0$ for an arbitrarily chosen time, and then go back using $u=-1/(2\kappa\varepsilon)$. This idea can be formulated as an exact mathematical statement, which is a part of the powerful \textit{saturation method} (see~\cite{JurdjevicKupka}).

Therefore, we can consider a new controllable system (recall that we set $n\equiv0$)
\begin{equation}
\label{eq:saturated_system}
	\dot\bbr =- \mathrm{Ad}_{\rme^{sf_1}}\big(\omega f_0(\bbr) + 2\kappa f_1(\bbr)u\big),
\end{equation}
where $(\mathrm{Ad}_\Phi f)(x) = \rmd\Phi^{-1}\circ f\circ \Phi = (\rmd\Phi(\Phi(x)))^{-1}[f(\Phi(x))]$ as usual, and $s$ is a new additional control. System (\ref{eq:saturated_system}) is usually called \textit{saturated system}. As we explain, the set of asymptotically reachable points for the new system coincides (up to some boundary points) with the set for the original system~(\ref{eq:main_control system}).

It is well known that if the right hand side of admissible velocities forms a non-convex or non-closed sets, one can take its closure and convex hull, and this procedure will not change the set of asymptotically reachable points. This idea can be also formulated as an exact mathematical statement called \textit{relaxation} of a control system (which is also a part of saturation method, see~\cite{JurdjevicKupka}).

Let us explain the relaxation procedure on the following example. Suppose that $\xi$ and $\eta$ are admissible velocities. Take an arbitrary $\lambda\in(0;1)$. Now we can plug $\xi$ for time $\lambda\varepsilon$, then plug $\eta$ for time $(1-\lambda)\varepsilon$, then plug $\xi$ again for time $\lambda\varepsilon$ and so on. As a result, we obtain a trajectory that moves approximately with the speed $\lambda\xi+(1-\lambda)\eta$. Hence, the hull is convex.

The last we can do with admissible velocities is to multiply them by arbitrary positive constant. Indeed, this will change only parametrization on the trajectory but not the trajectory itself.

Summarizing, the reachable sets for the system~(\ref{eq:main_control system}) with $n\equiv 0$ coincides up to some boundary points with the reachable sets of the extended system
\begin{equation}
\label{eq:relaxed_system}
	\dot\bbr \in \cl\mathrm{cone}\,\mathrm{conv}\left\{\mathrm{Ad}\,\rme^{sf_1}\big(\omega f_0(\bbr) + 2\kappa f_1(\bbr)u\big),\, \textrm{where}\, s\in\R,\ u\in\R\right\}.
\end{equation}
System (\ref{eq:relaxed_system}) is called \textit{relaxed system}.

Now we are ready to prove the inclusion $\cl C_+(\bbr^0)\supset \cl B_+(\bbr^0)$. Since $u\in\R$ can be taken arbitrary large, and the field $f_1$ is the rotation around the axis $O\bbr_x$, it seems natural to use cylindrical coordinates. Let
\[
	\bbr_x=z;
    \qquad \bbr_y=R\cos\theta;
    \qquad \bbr_z=R\sin\theta.
\]
Control system~(\ref{eq:main_control system}) takes the form
\begin{equation}
\label{eq:main_in_zRtheta}
	(\dot z, \dot R, \dot\theta)^T = \omega g_0(z,R,\theta) + 2\kappa g_1(z,R,\theta)u + \gamma g_2(z,R,\theta)n,
\end{equation}
where
\begin{eqnarray*}
	g_0(z,R,\theta) &=
	\left(\begin{array}{c}
		-R\cos\theta\\
		z\cos\theta\\
		-\frac{z}{R}\sin\theta
	\end{array}\right)
	-
	\frac\gamma\omega\left(\begin{array}{c}
		\frac12z\\
		\frac14R(3-\cos2\theta)\\
		\frac14\sin2\theta\\
	\end{array}\right)
	+
	\frac\gamma\omega\left(\begin{array}{c}
		0\\
		\sin\theta\\
		\frac1R \cos\theta
	\end{array}\right);\\
	g_1(z,R,\theta) &=
    \left(\begin{array}{c}
		0\\0\\1
	\end{array}\right);
	\qquad
	g_2(z,R,\theta) =
	- \left(\begin{array}{c}
		\frac12 z\\
		\frac14R(3-\cos2\theta)\\
		\frac14\sin2\theta
	\end{array}\right).
\end{eqnarray*}

It is important to note that $\rme^{sg_1}:(z,R,\theta)\mapsto(z,R,\theta+s)$. So, if we add new admissible velocities $\omega g_0(z,R,\varphi)$ for all $\varphi\in\R$, then as we explained this will not change the set of asymptotically reachable points either for pure coherent control or for both coherent and incoherent controls.

Now it is easy to see that
\[
	\frac12\Bigl(g_0(z,R,\theta) + g_0(z,R,\theta+\pi)\Bigr) = \frac\gamma\omega g_2(z,R,\theta).
\]
Indeed, if we change $\theta$ to $\theta+\pi$ in $g_0(z,R,\theta)$, then terms $\cos\theta$ and $\sin\theta$ change the sing, while terms $\cos2\theta$ and $\sin2\theta$ do not. 

So, we have proved that the vector $\frac\gamma\omega g_2(z,R,\theta)$ belongs to the convex hull of the right hand side of the saturated system (\ref{eq:saturated_system}) (written in $(z,R,\theta)$ coordinates). Therefore $f_2(\bbr)$ belongs to the right hand side of the relaxed system (\ref{eq:relaxed_system}). Remind that both saturation and relaxation procedures do not change the sets of asymptotically reachable points. Hence $\cl C_+(\bbr^0)\supset \cl B_+(\bbr^0)$, which completes the proof.

Thereby, if one excludes the incoherent control $n\ge 0$ from the system~(\ref{eq:main_control system}), then the reachable set becomes smaller, but only some of its boundary points are lost. At the same, the minimal motion time may become larger.

\begin{remark}
	The present paper is devoted to the analysis of the reachable sets, so we assume that $n\equiv0$ in what follows.
\end{remark}

\section{Asymptotically reachable points}\label{sec:asymptotically}

Recall that a point $\bbr^1$ is called asymptotically reachable from the point $\bbr^0$, if for any arbitrary small $\varepsilon>0$ there exists a trajectory of the control system that ends at a point in the distance at most $\varepsilon$ from $\bbr^1$. Hence, the set of asymptotically reachable points is $\cl C_+(\bbr^0)$.

The main result of this section is that most of the asymptotically reachable points are in fact exactly reachable. Precisely, 

\begin{thm}
	For any $\bbr^0$,
	\begin{equation}
	\label{eq:inclusions}
		\Int (\cl C_+(\bbr^0)) \subset C_+(\bbr^0) \subset \cl C_+(\bbr^0) = \cl(\Int C_+(\bbr^0)).
	\end{equation}
\end{thm}

In other words, we make two non-obvious statements on the structure of the set $\cl C_+(\bbr^0)$ of the asymptotically reachable points.
\begin{itemize}
	\item According to the first inclusion, if a neighborhood of a point $\bbr^1$ consists of points that are asymptotically reachable from $\bbr^0$, then $\bbr^1$ is in fact exactly reachable from $\bbr^0$.
	\item According to the last equality, the difference between the set of asymptotically reachable points and the interior of the set of exactly reachable points is rather small and coincides with the latter boundary:
	\[
		\cl C_+(\bbr^0)\setminus \Int C_+(\bbr^0) = \partial\, \Int C_+(\bbr^0).
	\]
\end{itemize}

These two facts eliminate some very strange inconvenient situations that may appear in general control systems (e.g.\ for some strange systems, it may happen that the set of exactly reachable points is dense, but has empty interior).

To show~(\ref{eq:inclusions}), we use a classical scheme based on Krener's theorem (see \cite{AgrachevSachkov}). This theorem works for control systems of full rank. Let us briefly remind what it means. By definition, the rank of the control systems $\dot\bbr = \omega f_0(\bbr) + 2\kappa f_1(\bbr)u$ at a point $\bbr$ is the dimension of the following linear subspace:
\[
	\mathrm{Lie}(f_0,f_1)(\bbr) = \mathrm{span}(f_0,f_1,[f_0,f_1],[f_0,[f_0,f_1]],[f_1,[f_0,f_1]],\ldots)(\bbr).
\]
We say that the system rank is full, if its rank at any $\bbr$ is equal to the state space dimension.

We claim that the control system $\dot\bbr = \omega f_0(\bbr) + 2\kappa f_1(\bbr)u$ is full rank. Indeed, denote $f_3=[f_0,f_1]$, $f_4=[f_0,f_3]$, $f_5=[f_1,f_3]$, $f_6=[f_1,f_5]$, and $f_7=(\mathrm{ad}\,f_1)^4f_0$. Direct computation gives
\begin{eqnarray*}
		f_3&=\left(\begin{array}{c}
			\bbr_z\\
			\frac\gamma\omega (1-\frac12\bbr_z)\\
			-\bbr_x - \frac\gamma{2\omega} \bbr_y
		\end{array}\right);\quad
		f_4=\left(\begin{array}{c}
			\frac\gamma\omega(\bbr_z-2)\\
			(1-\frac{\gamma^2}{4\omega^2})\bbr_z\\
			\frac\gamma\omega \bbr_x -  (1-\frac{\gamma^2}{4\omega^2})\bbr_y
		\end{array}\right);\\
		f_5&=\left(\begin{array}{c}
			-\bbr_y\\
			\bbr_x+\frac\gamma\omega \bbr_y\\
			\frac\gamma\omega(1-\bbr_z)
		\end{array}\right);\quad
		f_6=\left(\begin{array}{c}
			-\bbr_z\\
			\frac\gamma\omega(2\bbr_z-1)\\
			\bbr_x + 2\frac\gamma\omega \bbr_y
		\end{array}\right);\quad
		f_7=\left(\begin{array}{c}
			\bbr_y\\
			-\bbr_x+4\frac\gamma\omega \bbr_y\\
			\frac\gamma\omega (1-4\bbr_z)
		\end{array}\right).
\end{eqnarray*}

Since $\det(f_1,f_3,f_5)= (\bbr_y^2-\bbr_z^3+\bbr_z^2)\gamma/\omega$ and $\det(f_1,f_3,f_6)=3\bbr_y\bbr_z^2\gamma/\omega $, the system has rank $3$ at all points with $\bbr_y\ne 0$ or $\bbr_z\ne 0,1$. If $\bbr_y=\bbr_z=0$, then $\det(f_3,f_4,f_6)=4\gamma^3/\omega^3\ne 0$. Finally, if $\bbr_y=0$ and $\bbr_z=1$, then $\det(f_1,f_3,f_7)=-3\gamma/\omega\ne 0$. Hence the system $\dot\bbr=\omega f_0(\bbr)+2\kappa f_1(\bbr)u$ has full rank.

Let us now remind the statement of Krener's theorem (see~\cite{AgrachevSachkov}). According to this theorem, if a control system rank is full, then $\bbr^0\in\cl\Int C_+(\bbr^0)$ and $\bbr^0\in\cl\Int C_-(\bbr^0)$ for any point $\bbr^0$. At the first sight, this fact seems to be hardly applicable. Nonetheless, this theorem appears to be a very powerful tool in geometric control theory. 

Using this theorem we are able to show a very convenient fact on how to transform asymptotic reachability to the exact one. Let $\bbr^0$ and $\bbr^1$ be arbitrary points. We now show that if an interior point of the set of asymptotically reachable points from $\bbr^0$ is asymptotically reachable from $\bbr^1$ in backward time, then $\bbr^1$ is exactly reachable from $\bbr^0$:
\[
	\textrm{if}\qquad\Int\cl C_+(\bbr^0)\cap \cl C_-(\bbr^1)\ne \emptyset\qquad
	\textrm{then}\qquad \bbr^1\in C_+(\bbr^0).
\]

To show this, we are now going to apply many times the following simple fact: if $A\cap \cl B\ne \emptyset$ and the set $A$ is open, then $A\cap B\ne \emptyset$. First, we obtain $\Int\cl C_+(\bbr^0)\cap C_-(\bbr^1)\ne \emptyset$, as $\Int\cl C_+(\bbr^0)$ is open. We then take an arbitrary point $\bbr^2$ in the latter intersection. So, $\bbr^2\in\cl\Int C_-(\bbr^2)$ by Krener's theorem and hence, $\Int\cl C_+(\bbr^0)\cap\cl\Int C_-(\bbr^2)\ne\emptyset$. The first set in this intersection is open. Therefore, $\Int\cl C_+(\bbr^0)\cap\Int C_-(\bbr^2)\ne\emptyset$ and hence $\cl C_+(\bbr^0)\cap\Int C_-(\bbr^2)\ne\emptyset$. The second set in the latter intersection is open and we can use the same procedure again: $C_+(\bbr^0)\cap\Int C_-(\bbr^2)\ne\emptyset$ and $C_+(\bbr^0)\cap C_-(\bbr^2)\ne\emptyset$. Since $\bbr^2\in C_-(\bbr^1)$, we have $C_-(\bbr^2)\subset C_-(\bbr^1)$. Hence, $C_+(\bbr^0)\cap C_-(\bbr^1)\ne\emptyset$. So there exists a point $\bbr^3$ that is exactly reachable from $\bbr^0$ in forward time and from $\bbr^1$ in backward time. Therefore $\bbr^1$ is exactly reachable from $\bbr^0$ in forward time, $\bbr^1\in C_+(\bbr^0)$, as stated.

Now we are ready to show the main result of the section. The inclusion $C_+(\bbr^0) \subset \cl C_+(\bbr^0)$ is obvious. First, let us show that $\Int\cl C_+(\bbr^0) \subset C_+(\bbr^0)$. Let $\bbr^1\in\Int\cl C_+(\bbr^0)$. Then $\Int\cl C_+(\bbr^0)\cap C_-(\bbr^1)\ne \emptyset$ as $\bbr^1\in C_-(\bbr^1)$. Hence, $\bbr^1\in C_+(\bbr^0)$ as we show previously. Hence, we obtain $\Int\cl C_+(\bbr^0) \subset C_+(\bbr^0)$ as needed.

Second, let us show that $\cl C_+(\bbr^0) = \cl(\Int C_+(\bbr^0))$. The inclusion $\cl C_+(\bbr^0)\supset \cl(\Int C_+(\bbr^0))$ is obvious. The opposite inclusion also follows from Krener's theorem. Indeed, let $\bbr^1\in\cl C_+(\bbr^0)$, then the theorem implies $\bbr^1\in\cl\Int C_+(\bbr^1)$. Since $C_+(\bbr^1)\subset \cl C_+(\bbr^0)$, we obtain $\bbr^1\in\cl\Int\cl C_+(\bbr^0)$. On the one hand, we already know that $\Int\cl C_+(\bbr^0) \subset C_+(\bbr^0)$, but on the other hand, the set $\Int\cl C_+(\bbr^0)$ is open, which implies $\Int\cl C_+(\bbr^0) \subset \Int C_+(\bbr^0)$. Therefore, $\cl C_+(\bbr^0) \subset \cl(\Int C_+(\bbr^0))$ as needed. This finishes the proof.

\section{Size of the reachable set}\label{sec:size}

We now consider the point $(0,0,1)$ of the system. This point is the most important point by the following reason: if one leaves the system uncontrollable for a while (i.e. plug $u=n=0$), then any point in Bloch ball will tend to $(0,0,1)$ exponentially. In other words, staring the controllable process for the first time we can assume that $\bbr^0=(0,0,1)$. Hence the reachable sets of the point $(0,0,1)$ are indeed the most important ones. 

In the present section, we show that $C_+(0,0,1)$ is strictly smaller than the Bloch ball. In particular, we obtain that the gap size is $\delta\sim \gamma/\omega$. Precisely, we prove

\begin{figure}
	\begin{center}
    \includegraphics[width=0.5\textwidth]{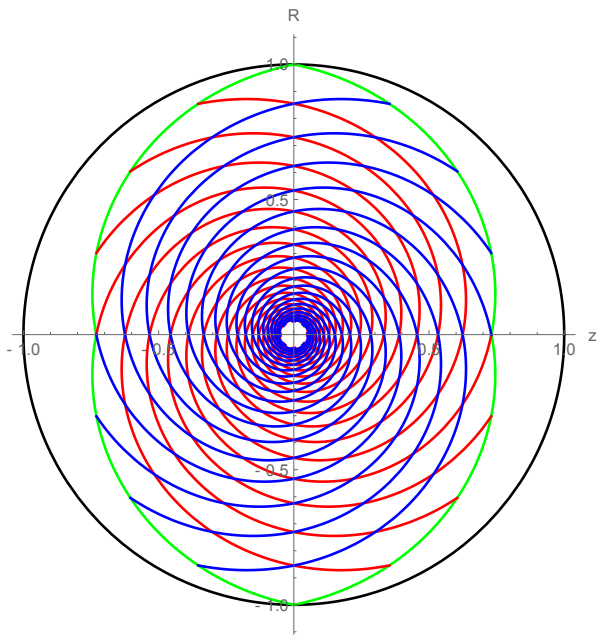}
    \end{center}
    \caption{The spiral grid in the polar coordinates.}
    \label{fig:spirals_grid}
\end{figure}

\begin{thm}
	\begin{itemize} For any $\gamma>0$ and $\omega>0$,
		\item The set $\cl C_+(0,0,1)$ is axisymmetrical w.r.t. $O\bbr_x$ and (obviously) contained in Bloch's ball. 
		\item The set $C_+(0,0,1)$ contains a 3D open set that consists of all possible rotations around $O\bbr_x$ of the following 2D set in the plane $O\bbr_x\bbr_y$ bounded by four parts of logarithmic spirals (see Figure~\ref{fig:spirals_grid})
			\[
				\bbr_x = \pm \rme^{-\gamma s/(2\omega)} \sin s; 
                 \qquad \bbr_y = \pm \rme^{-\gamma s/(2\omega)} \cos s; 
                 \qquad s\in[0;\frac{\pi}{2}].
			\]
		In particular, $C_+(0,0,1)$ contains a centered at the origin ball of radius $1-\frac{\pi}{4}\frac{\gamma}{\omega}$.
		\item On the other hand, any pure state (except for rotations of $(0,0,1)$ around $O\bbr_x$) has a neighborhood of size $\delta\sim\gamma/\omega$ that contain no points from $C_+(0,0,1)$. In particular, $\cl C_+(0,0,1)$ does not contain a centered at the origin ball of radius $1-\alpha\gamma/\omega$ for any $\alpha<\frac12(1+\frac{\gamma^2}{\omega^2})^{-1/2}$.
	\end{itemize}
\end{thm}

We have already shown in Section~\ref{sec:coherent_and_incoherent_controls} that the set $\cl C_+(0,0,1)$ is axis symmetric. The inclusion $C_+(0,0,1)\subset\{\bbr:|\bbr|\le 1\}$ is also obvious:
\[
\fl	\frac{\rmd}{\rmd t}\langle \bbr|\bbr\rangle = -\frac12\gamma(1+n)(\bbr_x^2+\bbr_y^2+2\bbr_z^2-2\bbr_z)=
	-\frac12\gamma(1+n)(\bbr_x^2+\bbr_y^2+\bbr_z^2+(1-\bbr_z)^2-1),
\]
which is non-positive if $|\bbr|=1$.

Let describe the main idea of the present section. We will work in cylindrical coordinates $z,R,\theta$ introduced in Section~\ref{sec:coherent_and_incoherent_controls}. Recall that from the asymptotic reachability point of view, since $u\in\R$ is unbounded, the function $\theta(t)$ in system~(\ref{eq:main_in_zRtheta}) can be chosen as close to a given arbitrary function as needed. So, let us consider an auxiliary system that is obtained by throwing away the equation on~$\dot\theta$ from system~(\ref{eq:main_in_zRtheta}):
\begin{equation}
\label{eq:main_two_dim_sys}
	\cases{
			\dot z = -\frac12\gamma z-\omega R \cos\theta;\\
			\dot R = \omega z\cos\theta -\frac14\gamma R(3-\cos2\theta) + \gamma\sin\theta.}
\end{equation}

In the auxiliary system~(\ref{eq:main_two_dim_sys}), $z$ and $R$ are new phase variables ($z^2+R^2\le 1$ and $R\ge 0$, but the non-negativity of $R$ is not essential as~(\ref{eq:main_two_dim_sys}) has an obvious symmetry $R\mapsto-R$, $\theta\mapsto\theta+\pi$), and $\theta\in\R/2\pi\Z$ is a new control. We claim that the auxiliary control system~(\ref{eq:main_two_dim_sys}) and the original system~(\ref{eq:main_in_zRtheta}) are connected in the following way: if a point is asymptotically reachable by the auxiliary control system, then its rotations around $O\bbr_x$ are asymptotically reachable in the original system and vise versa. In other words, $\cl C_+(\bbr^0)$ can be obtained as a union of all possible rotations around $O\bbr_x$ of the set of asymptotically reachable points from the corresponding point $(z^0,R^0)$ by the auxiliary control system. 

Let $A_+(z^0,R^0)$ denote the set of exactly reachable points from $(z^0,R^0)$ by the auxiliary control system~(\ref{eq:main_two_dim_sys}). So, we are going to show that
\begin{equation}
\label{eq:3d_and_2d_equal}
\fl	\cl C_+(\bbr^0) = \big\{
		\bbr: \bbr_x=z,\ \bbr_y=R\cos\theta,\ \bbr_z=R\sin\theta,\,\textrm{where}\,
		(z,R)\in \cl A_+(z^0,R^0),\ \theta\in\R
	\big\}
\end{equation}
where $z^0=\bbr^0_1\in\R$ and $R^0=((\bbr^0_2)^2+(\bbr^0_3)^2)^{1/2}\in\R$.

The inclusion $\subset$ is easy to show. Indeed, if $\bbr^1\in C_+(\bbr^0)$, then there exist controls $\hat u(t)$ and $\hat n(t)\equiv0$, and the corresponding trajectory $\bbr(t)$, $t\in[0;T]$, of the original system\footnote{Note that $\hat u\in L^1(0;T)$ and $\hat x\in W^1_1(0;T)$.}~(\ref{eq:main_control system}) such that $\hat \bbr(0)=\bbr^0$ and $\bbr(T)=\bbr^1$. Using coordinates $(z,R,\theta)$, we obtain a trajectory $(\hat z(t),\hat R(t),\hat\theta(t))$, which must be a solution to the auxiliary system\footnote{Note that $\hat z,\hat R,\hat\theta$ are in  $W^1_1(0;T)$.}~(\ref{eq:main_two_dim_sys}), where $\hat\theta(t)$ is free and can be considered as a control\footnote{Indeed, $W^1_1(0;T)\subset L^1(0;T)$.}. Hence, the end point $(z^1,R^1)$ of the constructed trajectory belongs to $A_+(z^0,R^0)$, i.e.
\[
\fl	C_+(\bbr^0) \subset \left\{
		\bbr: \bbr_x=z,\ \bbr_y=R\cos\theta,\ \bbr_z=R\sin\theta,\,
		\mbox{where}\, (z,R)\in A_+(z^0,R^0)
		\mbox{ and }\theta\in\R
	\right\}
\]
It remains to take closure of both sides to show the inclusion $\subset$ in~(\ref{eq:3d_and_2d_equal}).

We now prove the opposite inclusion~$\supset$. Let $(z^1,R^1)\in A_+(z^0,R^0)$. Let us fix an arbitrary angle $\theta^1$ and show that  $(z^1,R^1\cos\theta^1,R^1\sin\theta^1)\in\cl C_+(\bbr^0)$. We know that there exist a control $\hat\theta(t)$ and the corresponding trajectory $(\hat z(t),\hat R(t))$, $t\in[0;T]$, of the auxiliary system~(\ref{eq:main_two_dim_sys}) such that $\hat z(0)=z^0$, $\hat R(0)=R^0$, $\hat z(T)=z^1$, and $\hat R(T)=R^1$. Unfortunately, the control $\hat\theta\in L^1(0;T)$ may appear to be discontinuous, which is forbidden for the original system. Nonetheless, the space $C^\infty[0;T]$ is dense in $L^1(0;T)$, so let us choose a function $\tilde\theta\in C^\infty[0;T]$ that is close enough in $L^1$ norm to $\hat\theta$. Without loss of generality, we assume that $\tilde\theta(0)=\theta^0$ and $\tilde\theta(T)=\theta^1$. Now we consider a new trajectory $\tilde z(t),\tilde R(t)$ of the auxiliary system~(\ref{eq:main_two_dim_sys}) that is starting at the same point $(z^0,R^0)$, but uses new smooth control $\tilde\theta(t)$. Then the right end $(\tilde z(T),\tilde R(T))$ tends to $(z^1,R^1)$ as $\|\tilde\theta-\hat\theta\|_1\to0$. Moreover, new triplet $(\tilde z(t),\tilde R(t),\tilde\theta(t))$ satisfies equations on $\dot z$ and $\dot R$ of the original system~(\ref{eq:main_in_zRtheta}). It remains to chose the control $\tilde u(t)$ as the difference between $\frac{\rmd}{\rmd t}\tilde\theta$ and the right hand side of the corresponding equation in~(\ref{eq:main_in_zRtheta}). So any trajectory of the auxiliary system can be approximated by trajectories of the original system with an appropriate smooth control $\tilde u$. Hence,
\[
\fl \cl C_+(\bbr^0) \supset 
\left\{ \bbr: \bbr_x=z, \bbr_y=R\cos\theta, \bbr_z=R\sin\theta,\,
\mbox{where}\, (z,R)\in A_+(z^0,R^0)
\mbox{ and }\theta\in\R
\right\}.
\]
It remains to take closure of both side to prove the inclusion~$\supset$ in~(\ref{eq:3d_and_2d_equal}). So we have shown the first stated item.

Note that we have reduced the dimension of the system, but the cost is that the new control $\theta$ is nonlinear. Nonetheless, let us investigate the auxiliary system~(\ref{eq:main_two_dim_sys}) and its set $\cl A_+(0,1)$ of asymptotically reachable points. We are able to construct a lot of point that must belong to $\cl A_+(0,1)$. For example, we claim that $(0,-1)\in \cl A_+(0,1)$. Indeed, if one plugs $\theta\equiv -\pi/2$, then system~(\ref{eq:main_two_dim_sys}) becomes affine in $z$ and $R$ with a unique fixed point $(0,-1)$, and it is easy to check, that $(0,-1)$ is an asymptotically attracting point. Similarly, we obtain $(0,0)\in\cl A_+(0,1)$ by putting~$\theta\equiv0$.

Now consider more precisely trajectories of the auxiliary system. First, take $\theta\equiv0$ in the auxiliary control system~(\ref{eq:main_two_dim_sys}). In this case, the system is easy to solve: its solutions are logarithmic spirals the form $z+\rmi R=C\rme^{(-\gamma/2+\rmi\omega)t}$, $C\in\mathbb{C}$, which tend to $0$ and rotating clockwise. Second, take $\theta\equiv\pi$.  In this case, the system is also easy to solve: its solutions are logarithmic spirals of the form $z+\rmi R=C\rme^{(-\gamma/2-\rmi\omega)t}$, $C\in\mathbb{C}$, which tend to $0$ and are rotating counterclockwise. These two families of spirals form a grid on the disc $z^2+R^2\le 1$ shown on Fig.~\ref{fig:spirals_grid}, which is very convenient to use for controlling the auxiliary system. Let us take an arbitrary point $(z_1,R_1)$ in the disc that lies inside the domain bounded by four parts of green spirals:
\begin{equation}
\label{eq:spirals_t}
	z = \pm\rme^{-\frac12\gamma t} \sin \omega t; \qquad R = \pm \rme^{-\frac12\gamma t} \cos \omega t; \qquad t\in[0;\frac{\pi}{2\omega}].
\end{equation}
Using backward time motion, we start from $(z_1,R_1)$, and then using one of the grid logarithmic spirals, we go away from the origin until we reach one of the green spirals bounding the domain. Since each one of these 4 spirals start at $(0,1)$ or $(0,-1)$, we have $A_-(z_1,R_1)\cap A_+(0,1)\ne\emptyset$ or $A_-(z_1,R_1)\cap A_+(0,-1)\ne\emptyset$. In both cases, $(z_1,R_1)\in\cl A_+(0,1)$ as $(0,-1)\in \cl A_+(0,1)$. Note that during these motion, we sometimes use $R<0$, which is equivalent to the substitution $\tilde R=|R|$, $\tilde\theta=\theta + \frac12(1-\mathrm{sgn}\, R)\pi$.

Remind that if a point belongs to the interior of the set $\cl C_+(0,0,1)$ then it is in fact exactly reachable from $(0,0,1)$. So we obtain the second stated item by setting $s=\omega t$ in~(\ref{eq:spirals_t}). 

\begin{figure}\center
\includegraphics[width=0.5\textwidth]{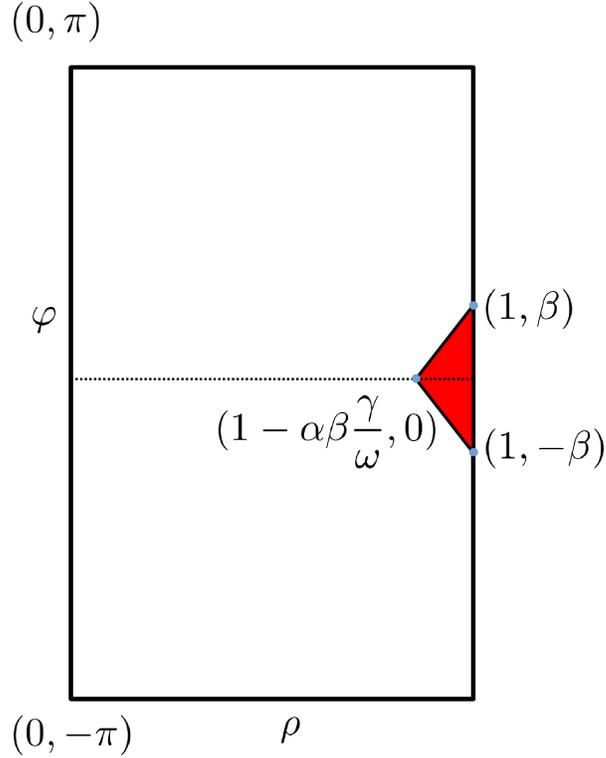}
\caption{The non-reachable triangle. Here $\alpha$ and $\beta$ are some sufficiently small parameters.}
\label{fig:cut_triangle}
\end{figure}

It remains to show that $\cl A_+(0,1)$ is strictly less than the unit disc $z^2+R^2\le 1$. Consider the polar coordinates on the disc (which corresponds to the spherical coordinates on Bloch's sphere). Denote $z=\rho\cos\varphi$ and $R=\rho\sin\varphi$. Then
\begin{eqnarray*}
		\dot\rho &= -\frac12\frac\gamma\omega \Big(\rho + \rho\sin^2\varphi\sin^2\theta-2\sin\varphi\sin\theta\Big);\\
		\dot\varphi &= -\cos\theta - \frac{1}{2\rho}\frac{\gamma}{\omega}\cos\varphi\sin\theta\Bigl(2-\rho\sin\varphi\sin\theta\Bigr).
\end{eqnarray*}

The variables $\rho$ and $\varphi$ belong to the rectangle $(\rho,\varphi)\in\Pi=[0;1]\times[-\pi;\pi]$, and pure states are given by $\rho=1$. Consider the previous system behavior in a neighborhood of the point $(\rho=1,\varphi=\varphi_0)$. Cut out from the rectangle $\Pi$ the triangle $\Delta$ bounded by three lines (see Fig.~\ref{fig:cut_triangle}): $\rho=1$ and $\rho=1\pm\alpha\frac\gamma\omega(\varphi-\varphi_0\mp\beta)$ where $\alpha>0$ and $\beta>0$ are some small parameters. Hence, $\Delta$ has vertices $(\rho=1-\alpha\beta\gamma/\omega,\varphi=\varphi_0)$ and $(\rho=1,\varphi=\varphi_0\pm\beta)$. We claim that if a trajectory starts at a point that does not belong to~$\Delta$, then it never intersects~$\Delta$ independently on the choice of control. 

To prove this, we show that all admissible velocities on the non-vertical triangle edges are directed outward of the triangle~$\Delta$. First, we compute outward normals to not vertical edges, which are $(-1,\pm\alpha\gamma/\omega)$. The dot product of these normals with the admissible velocities is
\[
	G_\pm=\left\langle (\dot\rho,\dot\varphi),\left(-1,\pm\alpha\frac\gamma\omega\right) \right\rangle.
\]
The functions $G_\pm(\rho,\varphi)$ have physical meaning only on the corresponding edges of $\Delta$, but one can formally compute their values also at the middle of the vertical edge:
\[
\fl	G_\pm\Big|_{\rho=1,\varphi=\varphi_0} =	\frac12\frac\gamma\omega\left(
		(1-\sin\varphi_0\sin\theta)^2 \mp \alpha \left(2\cos\theta + \frac\gamma\omega\cos\varphi_0\sin\theta(2-\sin\varphi_0\sin\theta)\right)
	\right).
\]
Functions $G_\pm$ are smooth analytic functions of $\alpha,\beta,\rho,\varphi$ and $\theta$. Also
\[
	G_{\pm}|_{\rho=1,\varphi=\varphi_0,\alpha=0}=\frac12\frac\gamma\omega\Bigl(1-\sin\varphi_0\sin\theta\Bigr)^2\ge \frac12\frac\gamma\omega\Bigl(1-|\sin\varphi_0|\Bigr)^2>0
\]
if $\varphi_0\ne\pm\frac\pi2$. Therefore, if $\alpha>0$ and $\beta>0$ are small enough, then $G_\pm$ are also positive on the non vertical edges\footnote{Here by the standard compactness argument we can use the same $\alpha$ and $\beta$ for all $\varphi_0$ outside any fixed neighborhood of $\pm\frac\pi2$.} of $\Delta$ for all $\theta$. Hence, any trajectory of the control system cannot intersect these edges from the outside. 

So any pure state $(\rho=1,\varphi\ne\frac\pi2)$ has a neighborhood of size $\delta\sim\gamma/\omega$ that contains no point reachable from $(0,1)$. In particular, we have proved that there is a gap in the Bloch ball that cannot be reached from the point $(0,0,1)$. It remains to find the gap size. We know that this gap is an axisymmetrical body (up to its boundary points) and lies outside from the centered at the origin ball of radius $1-\case{\pi}{4}\case{\gamma}{\omega}$ as was shown in the second item. So for $\varphi_0=0$ let us find a maximum possible value of $\alpha$ such that there exists some $\beta>0$ such that $G_\pm>0$ for all $\varphi\in[0;\beta]$ and $\rho=1\pm\alpha\frac\gamma\omega(\varphi\mp\beta)$. Let $\beta\to+0$ (and hence, $\varphi\to0$). In this case, $G_\pm\to \frac\gamma\omega[\frac12\mp\alpha(\cos\theta+\frac\gamma\omega\sin\theta)]$. So one can take any $\alpha<\frac12(1+\gamma^2/\omega^2)^{-1/2}$. This finishes the proof.

To estimate typical value of $\delta$, consider as an example 
calcium upper and lower levels $\rm 4^1 P$ and $\rm 4^1 S$ as two states
$|1\rangle$ and $|0\rangle$ of the two-level system whose all relevant parameters are known. For this system the
transition frequency is $\omega=4.5\times 10^{15}$~rad/s, the radiative
lifetime $t_{21}=4.5$~ns, so that $\gamma=1/t_{21}\approx
2.2\times 10^8$~s$^{-1}$, and the dipole moment $\kappa=2.4\times 10^{-29}$
C$\cdot$m~\cite{Hilborn}. In this case $\delta=\pi \gamma/4\omega\approx 6\cdot 10^{-9}$. For systems with larger decoherence rates, $\delta$ can be up to $10^{-2}$.

\begin{remark} 
	The analysis above is devoted to the most difficult and non-trivial case of only one coherent control in the Hamiltonian in the right hand side of Eq.~(\ref{f1}). For the case of two controls, the term $\kappa u(t)\sigma_x$ should be replaced by, e.g., $\kappa_1 u_1(t)\sigma_x+\kappa_2u_2(t)\sigma_y$. In this case, the system becomes completely controllable in the Bloch ball. Indeed, if a point on the distance $\rho$ from the origin belongs to the reachable set, then any point with the same distance to the origin belongs to the closure of the reachable set, since the controls $u_1$ and $u_2$ are unbounded. We already know, that for any $\rho<1$ there exists a reachable point (which can be attained just by using single control $u_1$). Hence any two interior points of the Bloch ball can be transferred in finite time one into the other by appropriate controls $u_1$ and $u_2.$
\end{remark}

\begin{figure}[t] 
		\includegraphics[width=1\linewidth]{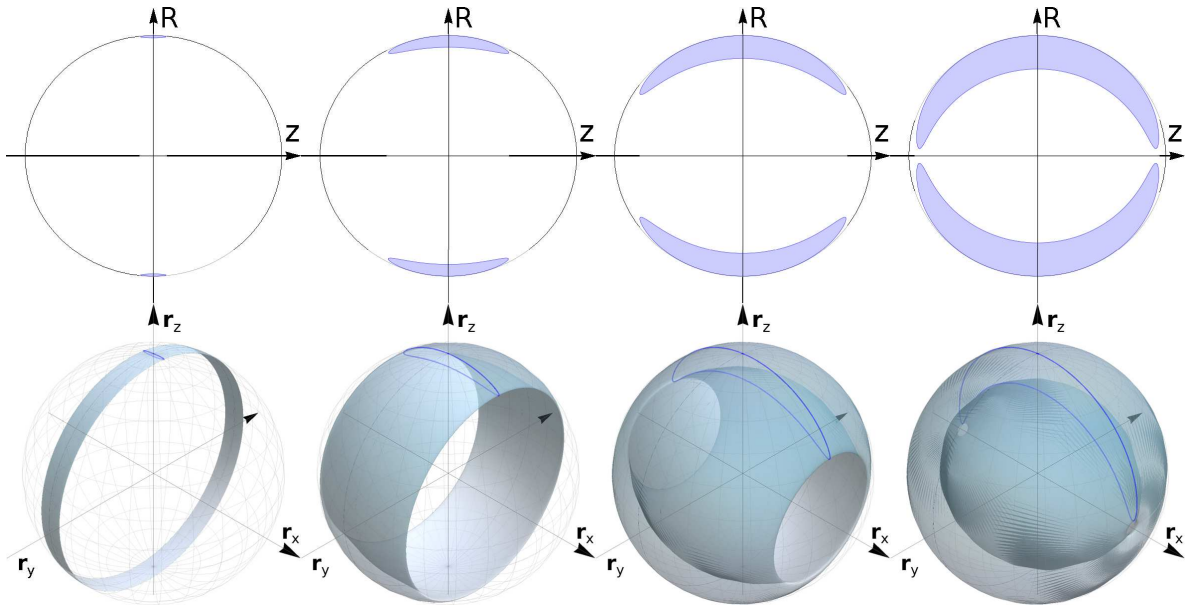}
		\caption{2D in the $(z,R)$ cylindrical coordinates (upper row) and 3D in the Bloch ball (bottom row) plots of the reachable sets for $\gamma/\omega=0.1$ and $\omega T=0.1$ (1st column), $\omega T=0.5$ (2nd column), $\omega T=1$ (3rd column), $\omega T=1.5$ (4th column).\label{figsets1}}
\end{figure}

\begin{figure}[h]
		\includegraphics[width=1\linewidth]{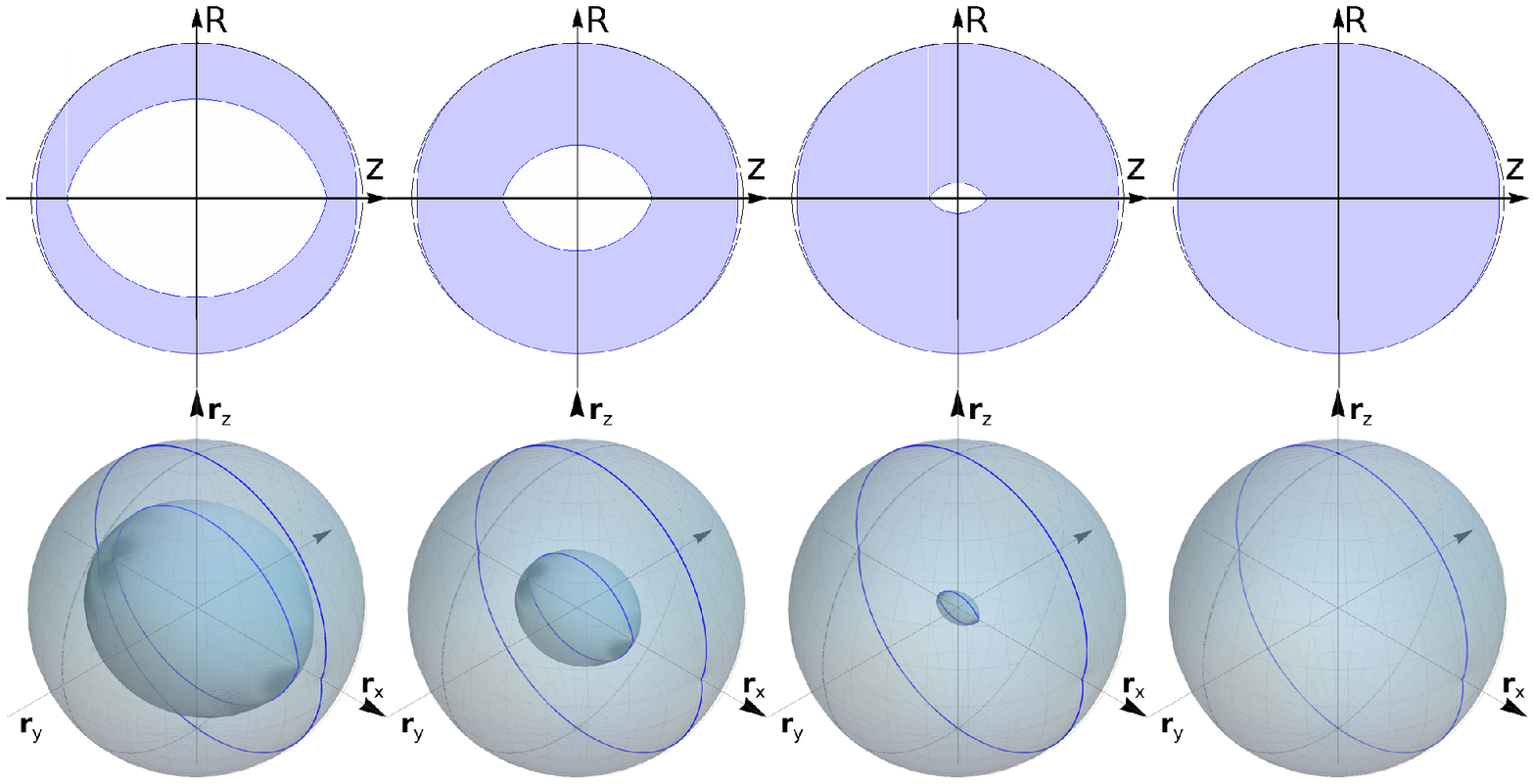}
		\caption{2D in the $(z,R)$ cylindrical coordinates (upper row) and 3D in the Bloch ball (bottom row) plots of the reachable sets for $\gamma/\omega=0.1$ and $\omega T=2$ (1st column), $\omega T=4$ (2nd column), $\omega T=6$ (3rd column), $\omega T=7$ (4th column).\label{figsets2}}
\end{figure}

\begin{figure}
\centering
\includegraphics[width=0.4\linewidth]{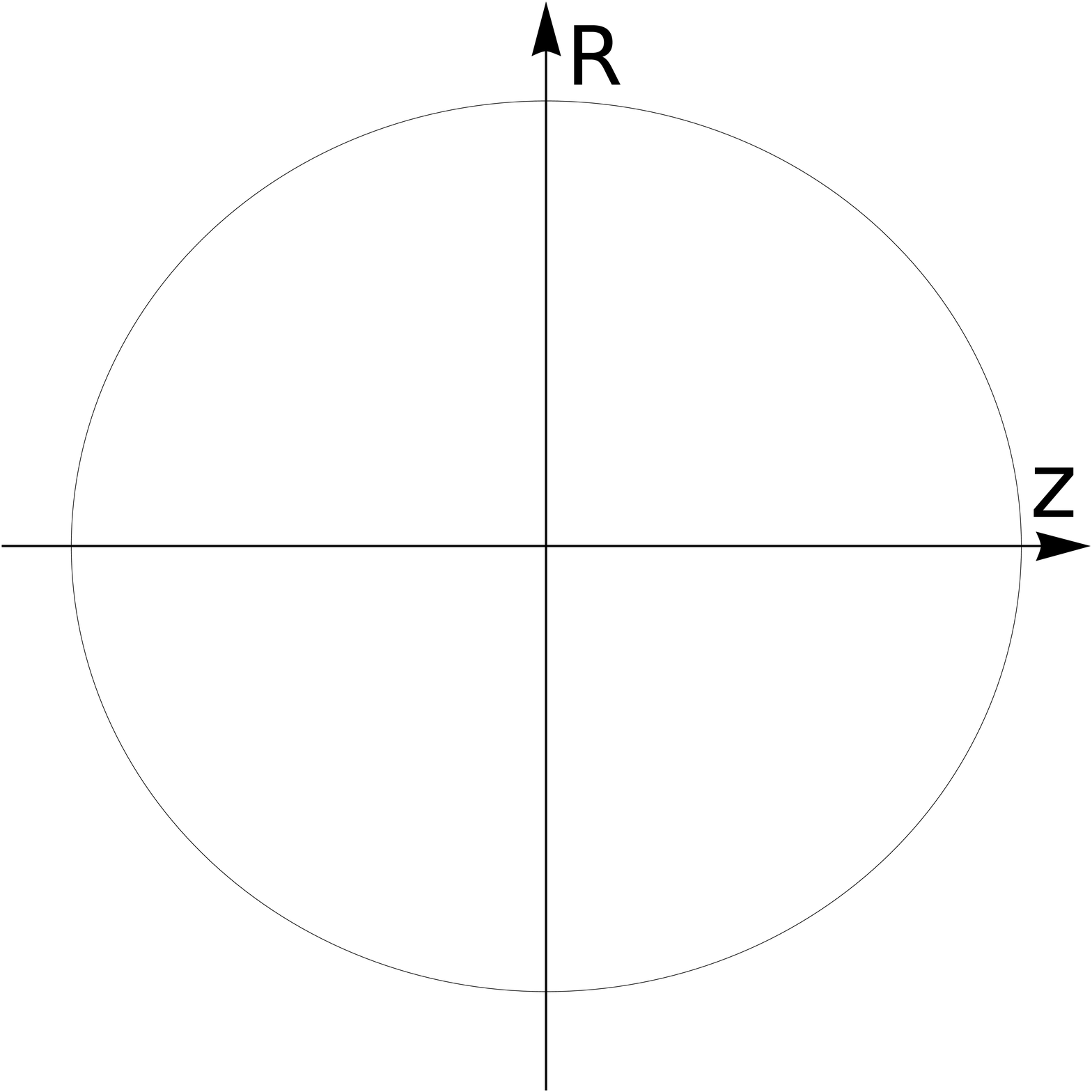}
\caption{Evolution of the reachable set as a function of $T$ (see the ancillary file ReachableSets.mp4 and supplementary
movie online at \href{https://stacks.iop.org/JPA/54/395304/mmedia}{https://stacks.iop.org/JPA/54/395304/mmedia}).
\label{figsetsmovie}}
\end{figure}

The reachable sets for the ground initial state $|0\rangle$ (which corresponds to the north pole of the Bloch sphere) for $\gamma/\omega=0.1$ were numerically computed and are shown for times $T=0.1, 0.5, 1, 1.5$ in the units of $1/\omega$ on Fig.~\ref{figsets1} and for $T=2, 4, 6, 7$ on Fig.~\ref{figsets2}. On each figure, upper rows show 2D plots on the disk in the cylindrical coordinates $(z,R)$ and bottom rows show 3D plots in the Bloch ball. Evolution of the reachable set as a function of $T$ is provided on Fig.~\ref{figsetsmovie} (movie is available in the online article). Note that we use only coherent control here. Using incoherent control can reduce the minimal time motion. Numerical computations rely on the ideas and formulas described in the next section.

\section{Fast numerical engineering of optimal coherent control}\label{sec:fast}

In this section, we present a method of designing optimal coherent control for time minimization problem $T\to\min$ for qubit control system~(\ref{eq:main_control system}). As we explained the optimal control $u(t)$ can be found as the difference between $\dot\theta$ and the r.h.s. of~(\ref{eq:main_in_zRtheta}), i.e.
\[
2\kappa u=\dot\theta +\omega\frac{z}{R} +\frac14\gamma \sin2\theta-\gamma\frac1R\cos\theta.
\]
where $\theta$ is a time optimal control in the auxiliary system~(\ref{eq:main_two_dim_sys}) (where that $R<0$ corresponds to substitution $\theta\mapsto\theta+\pi$). We use this method to numerically generate the reachable sets for $\gamma/\omega=0.1$ and for different times $T$. As described above, the results are shown on Fig.~\ref{figsets1} and Fig.~\ref{figsets2}; the upper 2D plots are in the cylindrical coordinates on the plane $(z,R)$ and the bottom 3D plots are obtained by rotation of the 2D sets around $O\bbr_x$ axis. Movie (in the online version of the paper) on Fig.~\ref{figsetsmovie} shows evolution of the reachable set as a function of $T$.

Let us rescale the time $\tau=t\omega$. System~(\ref{eq:main_two_dim_sys}) becomes
\begin{equation}
\label{eq:main_two_dim_sys_rescaled}
	\cases{
		z'_\tau = -\frac12\frac\gamma\omega z- R \cos\theta;\\
		R'_\tau =  z\cos\theta -\frac14\frac\gamma\omega R(3-\cos2\theta) + \gamma\sin\theta}.
\end{equation}

For the latter Pontryagin maximum principle gives
\[
	\mathcal{H}=-p\left(\frac12\frac\gamma\omega z+ R \cos\theta\right) + q\left( z\cos\theta -\frac14\frac\gamma\omega R(3-\cos2\theta) + \gamma\sin\theta\right),
\]
where $(p,q)$ are conjugate variables to $(z,R)$. So we have
\begin{equation}
\label{eq:PMP}
	\cases{
		p'_\tau = -\mathcal{H}'_z = \frac12\frac\gamma\omega p - q \cos\theta\\
		q'_\tau = -\mathcal{H}'_R =  p\cos\theta + \frac14\frac\gamma\omega q(3-\cos2\theta).
	}
\end{equation}

Pontryagin maximum principle states that for $t\in[0;T]$ optimal $\theta(t)$ solves maximization problem
\[
	\mathcal{H}(\theta)\to\max_{\theta\in[0;2\pi]}.
\]
Hence, $\theta$ that maximizes $\mathcal{H}$ becomes an (implicitly given) function $\theta(z,r,p,q)$. Unfortunately, numerical computation of a maximal point is a very time expensive procedure and it should be performed on each step of numerical ODE solving process. Hence, we suggest an alternative fast way to avoid this.

The idea is based on the fact that at any fixed point $(z,R)$, the r.h.s. of the rescaled auxiliary system~(\ref{eq:main_two_dim_sys_rescaled}) forms a strictly convex set for all $\theta\in[0;2\pi]$. Indeed, tangent vector $(\xi,\eta)$ to this set is
\[
	\xi=\frac{\rmd z'_\tau}{\rmd\theta} = R\sin\theta;
	\qquad
	\eta=\frac{\rmd R'_\tau}{\rmd\theta} = \frac\gamma\omega\big(\cos\theta-R\sin\theta \cos\theta\big)- z\sin\theta.
\]
It remains to note that vector $(\xi,\eta)$ rotates strictly clockwise as $\theta$ increases, since
\[
	\xi\eta'_\theta - \eta\xi'_\theta = \frac\gamma\omega  R(R\sin^3\theta-1) < 0.
\]
Hence $\theta(z,r,p,q)$ is a smooth function and its derivative can be easily found from linear on $\theta'_\tau$ equation $\mathcal{H}'_\theta=0$. Indeed, $\rmd \mathcal{H}'_\theta/\rmd\tau=0$ gives
\[
	\theta'_\tau = \frac18\frac\gamma\omega\left[
		\frac{(p R+q z)(5 \sin\theta+\sin3\theta) -8 p -4 \frac\gamma\omega  q \cos^3\theta}
			{(p R-q z) \cos\theta-\frac\gamma\omega  q (\sin\theta+R \cos2 \theta)}
		\right].
\]

Summarizing, together with~(\ref{eq:main_two_dim_sys_rescaled}) and~(\ref{eq:PMP}), we obtain an ODE system on $(z'_\tau,R'_\tau,p'_\tau,q'_\tau,\theta'_\tau)$, which can be numerically solved very fast. The last thing we need is to find initial values of variables. For simplicity, assume that we are starting from the attracting point\footnote{These initial values can be chosen arbitrary.}
\[
z(0)=0\quad\textrm{and}\quad R(0)=1.
\]
Conjugate variables can be rescaled by arbitrary positive multiplier, so without loss of generality we assume that $p(0)^2+q(0)^2=1$, i.e.
\[
	p(0)=\cos\psi_0\quad\textrm{and}\quad q(0)=\sin\psi_0,
\]
where $\psi_0$ determine optimal trajectory starting from $(z=0,R=1)$. Last initial value $\theta(0)=\theta_0$ can be found by solving simplified problem $\mathcal{H}(\theta)\to\max$ just once for $\tau=0$:
\[
	(\mathcal{H}'_\theta)\big|_{z=0,R=1} = \cos\psi^0\sin\theta^0-\frac\gamma\omega \sin \psi^0\cos\theta^0(\sin\theta^0-1)=0
\]
or
\[
	\cos\psi_0\sin\theta^0 = \frac\gamma\omega \sin\psi^0\cos\theta^0(\sin\theta^0-1).
\]
The latter equation has two solutions (one for maximal and another for minimal values of $\mathcal{H}$).

Now we give the last advise on reducing computational time. Note that system~(\ref{eq:main_two_dim_sys_rescaled}) depends only on one parameter $\gamma/\omega$. Suppose that $\gamma/\omega$ is given, and we have numerically found a number of optimal trajectories for a $\psi_0$ grid on $[0;2\pi]$. Then we can save the following table $(z^1,R^1)\mapsto(\psi^0,\theta^0)$ -- for each pair $(z^1,R^1)$ in a grid on the unit circle $z^2+R^2\le 1$ we save the initial values of conjugate variables and the initial angle $\theta_0$ of the corresponding optimal trajectory. Once the table is created, optimal trajectory from $(z=0,R=1)$ to any point $(z^1,R^1)$ can be found very fast, since we can immediately extract optimal initial values $(\psi_0,\theta_0)$ from the table.

\section{Conclusions}\label{sec:conclusions}

This work provides an explicit analytical study of reachable sets for an open two-level quantum system driven by coherent and incoherent controls. First, it is shown that for the two-level case the presence of incoherent control does not affect the reachable set (while incoherent control may affect the time necessary to reach particular state). Second, the reachable set is explicitly described in the Bloch ball. For the case of one coherent control it is shown that most points in the Bloch ball can be exactly reached, except of two lacunae of the size $\delta\sim \gamma/\omega$, where $\gamma$ is the decoherence rate and $\omega$ is the transition frequency. Thus any point in the Bloch ball can be achieved with precision $\delta$. Typical values are $\delta\lesssim10^{-3}$ that implies high accuracy of achieving any density matrix. For two coherent controls, the system is shown to become completely controllable in the set of all density matrices. Third, the reachable set as a function of the final time is explicitly described and shown to exhibit a non-trivial structure.

\ack

This work was funded by Russian Federation represented by the Ministry of Science and Higher Education (grant number 075-15-2020-788).


\bigskip

\begin{thebibliography}{99}

\bibitem{Glaser2015}
S.~J.~Glaser, U.~Boscain, T.~Calarco, C.~P.~Koch, W.~K\"{o}ckenberger, 
R.~Kosloff, I.~Kuprov, B.~Luy, S.~Schirmer, T.~Schulte-Herbr\"{u}ggen,
D.~Sugny, F.~K.~Wilhelm, Training Schr\"{o}dinger's cat: quantum optimal control.
Strategic report on current status, visions and goals for research in Europe',
{\it Eur. Phys. J. D}, {\bf 69}:12, 279 (2015).

\bibitem{ButkovskyBook1994}
A.~G.~Butkovskiy, Y.~I.~Samoilenko, {\it Control of Quantum-Mechanical Processes and Systems} (Nauka Publ., Moscow, 1984) (In Russian.) English transl.: A.G.~Butkovskiy, Yu.I.~Samoilenko,  {\it Control of quantum-mechanical processes and systems} (Kluwer Acad. Publ., Dordrecht, 1990).

\bibitem{ShapiroBrumerBook2003}
M. Shapiro, P. Brumer, {\it Principles of the Quantum Control of Molecular Processes}
(John Wiley \& Sons, Inc., Hoboken, 2003).

\bibitem{TannorBook2007}
D.~J.~Tannor, {\it Introduction to Quantum Mechanics: A Time Dependent Perspective}
(Univ. Science Books, Sausilito, CA, 2007).

\bibitem{AlessandroBook2007}
D.~D'Alessandro, {\it Introduction to Quantum Control and Dynamics} 
(CRC Press, Boca Raton, 2007).

\bibitem{LetokhovBook2007}
V. Letokhov, {\it Laser Control of Atoms and Molecules} 
(Oxford Univ. Press, 2007).

\bibitem{FradkovBook2007}
A.~L. Fradkov, {\it Cybernetical Physics. From Control of Chaos to Quantum Control}
(Springer, New York, 2007).

\bibitem{BrifChakrabartiRabitz2010}
C. Brif, R. Chakrabarti, H. Rabitz, Control of quantum phenomena: 
past, present and future, New J.~Phys. \textbf{12}:7, 075008 (2010).

\bibitem{WisemanMilburnBook2010}
H.~M.~Wiseman, G.~J.~Milburn, {\it Quantum Measurement and Control}
(Cambridge Univ. Press, Cambridge, 2010).

\bibitem{DongPetersen2010}
D.~Dong and I.R.~Petersen, Quantum control theory and applications: a survey,
{\it IET Control Theory and Applications}, {\bf 4}:12, 2651--2671 (2010). 

\bibitem{Moore2011} 
K.~W. Moore, A. Pechen, X.-J. Feng, J. Dominy, V.~J. Beltrani, H.~Rabitz, Why is chemical synthesis and property optimization easier than expected?, Physical Chemistry Chemical Physics, {\bf 13}:21, 10048--10070 (2011). 

\bibitem{ZagoskinBook2011}
A.M. Zagoskin, {\it Quantum Engineering. Theory and Design of Quantum Coherent Structures} (Cambridge Univ. Press, Cambridge, 2011).

\bibitem{CPKoch2016OpenQS} 
C.~P. Koch, Controlling open quantum systems: Tools, achievements, 
and limitations, J.~Phys.: Condens. Matter. \textbf{28}:21, 213001 (2016). 

\bibitem{Stefanatos2020} D. Stefanatos and E. Paspalakis, A shortcut tour of quantum control methods for modern quantum technologies, EPL (Europhysics Letters), \textbf{132}, 60001 (2020).

\bibitem{PechenRabitz2006} A. Pechen, H. Rabitz, Teaching the environment to control  quantum systems, Phys. Rev.~A. {\bf 73}:6, 062102 (2006).

\bibitem{PechenIlinShuangRabitz2006} A. Pechen, N. Il'in, F. Shuang, H. Rabitz, Quantum control by  von Neumann measurements, Phys. Rev. A. {\bf 74}, 052102 (2006).

\bibitem{ShuangPechenHoRabitz2007} 
F.~Shuang, A.~Pechen, T.-S.~Ho, H.~Rabitz, Observation-assisted optimal 
control of quantum dynamics, J.~Chem. Phys., \textbf{126} (13), 134303 (2007).

\bibitem{ShuangZhouPechenWuShirRabitz2008} 
F.~Shuang, M.~Zhou, A.~Pechen, R.~Wu, O.~M.~Shir, H.~Rabitz, Control of quantum dynamics by optimized measurements, Phys. Rev.~A, \textbf{78} (6), 063422 (2008).

\bibitem{PechenTrushechkin2015} 
A.~Pechen and A.~Trushechkin, Measurement-assisted Landau-Zener 
transitions, Phys. Rev.~A, \textbf{91}, 052316 (2015).

\bibitem{Pechen2011} A. Pechen, Engineering arbitrary pure and mixed quantum states, Phys. Rev.~A, {\bf 84}, 042106 (2011).

\bibitem{Wu2007}
R.~Wu, A.~Pechen, C.~Brif, H.~Rabitz, Controllability of open quantum systems with Kraus-map dynamics, J.~Phys.~A: Math.~Theor., \textbf{40}:21, 5681--5693 (2007).

\bibitem{MorzhinIJTP2019} O.V. Morzhin and A.N. Pechen, Minimal time generation of density matrices for a two-level quantum system driven by coherent and incoherent controls, Internat. J. Theoret. Phys., {\bf 60}, 576--584  (2021; Published Online 2019). 

\bibitem{JurdjevicKupka} V. Jurdjevic and I. Kupka, Control systems subordinated to a group action: accessibility, Journal of Differential Equations, {\bf 39}, 186--211 (1981).

\bibitem{AgrachevSachkov} A. Agrachev, Yu. Sachkov, {\it Control Theory from the Geometric Viewpoint} (Springer-Verlag Berlin Heidelberg, 2004).

\bibitem{arXiv:2103.10898} 
M.~Rademacher, M.~Konopik, M.~Debiossac, D.~Grass, E.~Lutz, and  N.~Kiesel,  Nonequilibrium control of thermal and mechanical changes in a levitated system,  
arXiv:2103.10898. 

\bibitem{Huang1983} G. M. Huang and T. J. Tarn, On the controllability of quantum‐mechanical systems, Journal of Mathematical Physics, {\bf 24}, 2608 (1983).

\bibitem{Tarn1984} T.J. Tarn, J.W. Clark, G.M. Huang, Analytic controllability of quantum-mechanical systems. In: Fuhrmann P.A. (eds) Mathematical Theory of Networks and Systems. Lecture Notes in Control and Information Sciences, vol 58. Springer, Berlin, Heidelberg  (1984). 

\bibitem{Turinici2001}
G. Turinici and H. Rabitz, Quantum wavefunction controllability, Chem. Phys., {\bf 267}, 1--9 (2001).

\bibitem{Albertini2001}
F. Albertini and D. D’Alessandro, Notions of controllability for quantum-mechanical systems, arXiv quant-ph/0106128 (2001).

\bibitem{Fu2001}
H. Fu, S. G. Schirmer, A. I. Solomon, Complete controllability of finite-level quantum systems,  J. Phys. A: Math. Gen., {\bf 34}, 1679 (2001).

\bibitem{Schirmer2001}
S. G. Schirmer, H. Fu, A. I. Solomon, Complete controllability of quantum systems, Phys. Rev. A, {\bf 63}, 063410 (2001).

\bibitem{Schirmer2002}
S. G. Schirmer, A. I. Solomon, J. V. Leahy, Criteria for dynamical reachability of quantum
states, J. Phys. A: Math. Gen., {\bf 35}, 8551--8562 (2002).

\bibitem{Altafini2002}
C. Altafini, Controllability of quantum mechanical systems by root space decompositions of $\mathrm{su}(n)$, J. Math. Phys., {\bf 43}, 2051--2062 (2002).

\bibitem{Polack2009} T. Polack, H. Suchowski, D. J. Tannor, Uncontrollable quantum systems: A classification scheme based on Lie subalgebras, Phys. Rev. A {\bf 79}, 053403 (2009).

\bibitem{Boscain2015} U. Boscain, J.P. Gauthier, F. Rossi, M. Sigalotti, Approximate controllability, exact controllability, and conical eigenvalue intersections for quantum mechanical systems, Commun. Math. Phys., {\bf 33}, 1225--1239 (2015). 

\bibitem{Altafini1} C. Altafini, Controllability properties for finite dimensional quantum Markovian master equations, J. Math. Phys., {\bf 44}(6), 2357--2372 (2003).

\bibitem{Altafini2} C. Altafini, Controllability of open quantum systems: The two level case, Published in 2003 IEEE International Workshop on Workload Characterization (IEEE Cat. No.03EX775) {http://dx.doi.org/10.1109/PHYCON.2003.1236992}

\bibitem{SugnyPRA2007} D. Sugny, C. Kontz, H.~R. Jauslin, Time-optimal control of a two-level dissipative quantum system, Phys. Rev. A, {\bf 76}, 023419 (2007).

\bibitem{BonnardSIAM2009}  B. Bonnard and D. Sugny, Time-minimal control of dissipative two-level quantum systems:
The Integrable case, SIAM J. Control Optim. {\bf 48}, 1289--1308 (2009).

\bibitem{BonnardIEEE2009} B. Bonnard, M. Chyba, D. Sugny, Time-minimal control of dissipative two-level quantum systems: The generic case, IEEE Transactions on Automatic Control, {\bf 54}, 2598--2610 (2009).

\bibitem{StefanatosPRA2009} D. Stefanatos, Optimal design of minimum-energy pulses for Bloch equations in the case of dominant transverse relaxation, Phys. Rev. A, {\bf 80}, 045401 (2009).

\bibitem{StefanatosSCL2010} D. Stefanatos and J.-S. Li, Constrained minimum-energy optimal control of the dissipative Bloch equations, Systems \& Control Letters, {\bf 59}, 601--607 (2010).

\bibitem{MorzhinLJM2020} O. Morzhin and A. Pechen, Machine learning for finding suboptimal final times and coherent and incoherent controls for an open two-level quantum system, Lobachevskii J. Math., {\bf 41}:12, 2353--2368 (2020).

\bibitem{MorzhinAIP2021}
O. Morzhin and A. Pechen, Numerical estimation of reachable and controllability sets for a two-level open quantum system driven by coherent and incoherent controls, AIP Conf. Proc., {\bf 2362}, 060003 (2021).

\bibitem{PechenPRA2012} A. Pechen and N. Il'in, Trap-free manipulation in the Landau-Zener system, Phys. Rev. A, {\bf 86}, 052117 (2012).

\bibitem{PechenJPA2017}
A. Pechen and N. Ilin, Control landscape for ultrafast manipulation by a qubit, J. Phys. A: Math. Theor., {\bf 50}:7, 75301 (2017).

\bibitem{VolkovJPA2021} B. O. Volkov, O. V. Morzhin, A. N. Pechen, Quantum control landscape for ultrafast generation of single-qubit phase shift quantum gates, J. Phys. A: Math. Theor., {\bf 54}:21, 215303 (2021).

\bibitem{Hilborn} R. C. Hilborn, American J. Phys. {\bf 50}, 982 (1982).

\end{thebibliography}
\end{document}